\documentclass{amsart}

\usepackage{amsmath}
\usepackage{amsfonts}
\usepackage{amsthm}

\DeclareMathOperator{\curl}{curl}

\DeclareMathOperator{\diverg}{div}
\DeclareMathOperator{\id}{id}

\DeclareMathOperator{\ran}{ran}

\newtheorem{lemma}{Lemma}[section]
\newtheorem{theorem}[lemma]{Theorem}

\newtheorem{corollary}[lemma]{Corollary}
\theoremstyle{definition}
\newtheorem{definition}[lemma]{Definition}
\theoremstyle{definition}
\newtheorem{remark}[lemma]{Remark}
\theoremstyle{definition}

{\catcode `\@=11 \global\let\AddToReset=\@addtoreset}
\AddToReset{equation}{section}

\newcommand{\ie}{{\sl i.e.\/ }}
\newcommand{\cf}{{\sl cf.\/ }}
\newcommand{\eg}{{\sl e.g.\/}}

\newcommand{\newpar}{\par}\parindent =0pt\parskip=3pt\textheight = 615pt

\newcommand{\Id}[1]{{\rm I\kern-2pt I_{#1}}}
\renewcommand{\hbar}{{\displaystyle\bar{\phantom{x}}\kern-6pt h}}

\numberwithin{equation}{section}

\begin{document}
\thispagestyle{empty}
\begin{center}
\fontsize{20}{22} \selectfont \textbf{Semiclassical Asymptotics
for the Maxwell - Dirac System}
\end{center}
\vskip 1 cm
\begin{center}
\fontsize{12}{12} \selectfont C. Sparber \footnote{e-mail:
christof.sparber@univie.ac.at} and P. Markowich \footnote{e-mail:
peter.markowich@univie.ac.at } 
\end{center}
\begin{center}
\vskip 0.5cm
Wolfgang Pauli Institute Vienna and\\
Department of Mathematics, University of Vienna,\\
Strudlhofgasse 4, A-1090 Vienna,\\
Austria \\
\end{center}
\vskip 1 cm
\begin{center}
\textbf{Abstract}
\end{center}
\emph{We study the coupled system of Maxwell and Dirac equations from a semiclassical point of view.
A rigorous nonlinear WKB-analysis, locally in time, for solutions of (critical) order $O(\sqrt{\varepsilon })$ is performed,
where the small semiclassical parameter $\varepsilon $ denotes the microscopic/macroscopic scale ratio.}
\newpar
\newpar
\noindent \textbf{AMS (2000) subject classification:} 81Q20, 35B25, 35B40, 35L60
\vskip 1cm


\section{Introduction and Scaling}

The \emph{Maxwell-Dirac system} (MD) is fundamental in the relativistic description of spin $1/2$ particles.
It represents the time-evolution of fast (relativistic) electrons and positrons within external and \emph{self-consistent}
generated electromagnetic fields:
\begin{equation}\label{dm0}
\left \{
\begin{split}
& i \hbar \partial_ s  \Psi   = \sum_{k=1}^3 \alpha^k   (-i\hbar c\partial_k -e (A_k+A_k^{ext})) \Psi + e(V+V^{ext})\Psi
+mc^2 \beta \Psi ,\\
& \left(\frac{1}{c^{2}} \partial^2_ s - \Delta \right) V = \frac{1}{\epsilon_0} \rho,
\quad \left(\frac{1}{c^{2}} \partial^2_ s - \Delta\right) A = \frac{1}{c\epsilon_0} j,
\end{split}
\right.
\end{equation}
where the \emph{particle-} and \emph{current-densities} are defined by:
\begin{equation}
\rho:= e|\Psi|^2, \quad j_k:= e\langle \Psi , \alpha^k \Psi \rangle, \quad \ k=1,2,3.
\end{equation}
Here, $\Psi = \Psi (s,y)\in \mathbb C^4$ is the $4$-vector of the \emph{spinor field}, normalized s.t.
\begin{equation}\label{nor}
\int_{\mathbb R^3} |\Psi(s,y) |^2 dy =1,
\end{equation}
with $s$, $y\equiv (y_1,y_2,y_3)$, denoting the time - resp. spatial coordinates
in \emph{Minkowski space}.
Further, $V^{(ext)} =V^{(ext)}(s,y)\in \mathbb C$ is the \emph{self-consistent} resp. \emph{external electric potential} and
$A^{(ext)}_k =A^{(ext)}_k(t,x)\in \mathbb C$, the corresponding \emph{magnetic potential}, with $A=(A_1,A_2,A_3)$. In the following,
the usual scalar-product for vectors $X, Y\in \mathbb C^4$ will be denoted by $\langle X ,Y \rangle$ and we shall
also write $|X|^2:=\langle X , X \rangle$.
The so-called \emph{Dirac matrices} $\beta, \alpha^k$, $k=1,2,3$, are explicitly given by:
\begin{equation}
\beta:=
\begin{pmatrix}
\Id{2} & 0\\
0 & -\Id{2}
\end{pmatrix}
,\quad
\alpha^k:=
\begin{pmatrix}
0 & \sigma ^k\\
\sigma ^k & 0
\end{pmatrix},
\end{equation}
with $\Id{2}$, the $2\times 2$ identity matrix and $\sigma ^k$ the $2\times 2$ \emph{Pauli matrices}, \ie
 \begin{equation}
\sigma  ^1:=
\begin{pmatrix}
0 & 1\\
1 & 0
\end{pmatrix}
,\quad
\sigma  ^2:=
\begin{pmatrix}
0 & -i\\
i & 0
\end{pmatrix}
,\quad
\sigma  ^3:=
\begin{pmatrix}
1 & 0\\
0 & -1
\end{pmatrix}.
\end{equation}
Hence, $\alpha^k, \beta$ are \emph{hermitian} and moreover one easily checks that the
following identities hold for $ k,l=1,2,3$:
\begin{equation}
\left \{
\begin{split}
\alpha^k \alpha^l+ \alpha^l \alpha^k= & \ 2\delta _{kl},\quad \\
\alpha^k\beta +\beta  \alpha^k= & \ 0.
\end{split}
\right.
\end{equation}
Finally, the appearing physical constants are the normalized
Planck constant $\hbar$, the speed of light $c$, the permittivity
of the vacuum $\epsilon_0$, the particle mass $m$ and the charge
$e$.
\newpar
Additionally to (\ref{dm0}), we impose the \emph{Lorentz gauge condition}
\begin{equation}
\frac{1}{c} \partial _s V(s,y) + \diverg A(s,y)=0,
\end{equation}
for the initial potentials $V(0,y)$ and $A(0,y)$. The gauge is henceforth conserved during the time-evolution.
It ensures that the corresponding \emph{electromagnetic fields} $E$, $B$ are uniquely determined by
\begin{equation}
E(s,y):=- \frac{1}{c} \partial _s A(s,y)-\nabla V(s,y), \quad B(s,y):=\curl A(s,y).
\end{equation}
The MD equations are the underlying field equations of relativistic \emph{quantum electro-dynamics}, \cf \cite{Sc},
where one considers the system within the formalism of \emph{second quantization}. Nevertheless,
in order to obtain a deeper understanding for the interaction of matter and radiation, there is a growing
interest in the MD system also  for classical fields, since one can expect at least qualitative results, \cf \cite{EsSe}.
\newpar
From the mathematical point of view, the strongly nonlinear MD system poses a hard problem in the study of PDE's
arising from quantum physics. Well posedness and existence of solutions on all of $\mathbb R^3_y$
but only locally in time, has been proved almost forty years ago in \cite{Gro}. On the other hand,
only partial results (\ie for small initial data)
have been obtained in the quest of global-in-time solutions, \cite{Ch}, \cite{G}, \cite{FST}, let alone
the study of other qualitative features of this system.
\newpar
In this paper, we shall analyze the MD system in a \emph{semiclassical} regime. To do so, we
first rewrite the equations such that there remains only one (positive) \emph{dimensionless} parameter
\begin{equation}
\delta = \frac{\hbar c \varepsilon_0 } {e^{2}},
\end{equation}
which is obtained by replacing $y \rightarrow y / \bar y$, $s
\rightarrow s / \bar s$ and $\Psi(s,y) \rightarrow \bar y ^{-3/2}
\Psi(s/ \bar s,y/ \bar y) $, in order to maintain the
normalization condition (\ref{nor}), with
\begin{equation}
c \bar s = \bar y \ \mbox{ and } \ \bar y= \frac{e^2}{mc^2 \varepsilon_0}.
\end{equation}
Here, we also replaced both, the external and the self-consistent potentials, by
$A^{(ext)}(s,y) \rightarrow \lambda A^{(ext)} (s/ \bar s,y/ \bar y)$ and
$V^{(ext)}(s,y) \rightarrow \kappa V^{(ext)} (s/ \bar s,y/ \bar y)$, with $\kappa = c\lambda$ and $\lambda = mc/e$.
We assume for the following that $A^{ext}, V^{ext}$ are of $O(1)$ in these units.
In summary, we obtain the MD system in dimensionless form:
\begin{equation}\label{dmd}
\left \{
\begin{split}
& i \delta \partial_ s  \Psi   = \sum_{k=1}^3 \alpha^k   (-i \delta \partial_k - (A_k+A_k^{ext}))\Psi + (V+V^{ext})\Psi
+ \beta \Psi ,\\
& \ \Box V =  \rho,
\quad \Box A =  j,
\end{split}
\right.
\end{equation}
where from now on $\Box:=\partial ^2_s-\Delta$. In (\ref{dmd}), $s,y$ represent the \emph{microscopic}
time and length scales.
Note that in general, $\delta$ can not be considered as a \emph{small} parameter, for example $\delta \approx 13$,
in the case of electrons. Hence, semiclassical asymptotics in $\delta$,
\ie on (\ref{dmd}) directly, only make sense for highly charged and consequently heavy particles.
\newpar
Therefore, we need to rescale the system (\ref{dmd}) such that the
time-evolution can be considered semiclassical, independent of
the precise physical properties of the particles. We can suppose
that the given external electromagnetic potentials are slowly
varying w.r.t to the microscopic scales, \ie $V^{ext}=V^{ext} (s
\varepsilon / \delta , y \varepsilon / \delta )$ and likewise
$A^{ext}=A^{ext}(s \varepsilon / \delta , y \varepsilon / \delta
)$, where from now on $\varepsilon$ denotes the small
\emph{semiclassical parameter}. Here, the $\delta$ is included in
the scaling in order to eliminate it from the resulting equation.
Hence, observing the evolution on macroscopic scales we are lead
to:
\begin{equation}
y = \frac{\delta }{\varepsilon} \, x, \quad s = \frac{\delta }{\varepsilon}\, t.
\end{equation}
Moreover, we want that the coefficients of all nonlinearities to be $O(1)$, \ie they should \emph{not} carry a positive
power of $\varepsilon$.
It turns out that there exists solutions $\psi^\varepsilon$, which obey this requirement. If we set
\begin{equation}
\frac{\delta }{ \varepsilon} \Psi (s,y) = \psi^\varepsilon (t,x).
\end{equation}
then, the normalization condition for (\ref{nor}) gives
\begin{align}
\int_{\mathbb R^3} |\psi^\varepsilon (t,x)|^2 dx  =
\frac{\varepsilon}{\delta} \int_{\mathbb R^3} |\Psi(s,y) |^2 dx = \frac{\varepsilon}{\delta}.
\end{align}
This implies that we need to look for solutions $\psi^\varepsilon$ s.t.
\begin{align}
\psi^\varepsilon \sim O(\sqrt{\varepsilon}),
\end{align}
assuming, as mentioned above, that $\delta \sim O(1)$. We therefore end up with the following
\emph{semiclassical scaled} MD system:
\begin{align}
\label{dm1} & i \varepsilon  \partial_ t  \psi ^\varepsilon  = \sum_{k=1}^3\alpha^k   (-i\varepsilon  \partial_k -
(A_k^{\varepsilon}+A^{ext})) \psi^\varepsilon +\beta \psi ^\varepsilon + (V^\varepsilon+V^{ext})\psi ^\varepsilon ,\\
\label{dm2} &\Box V^\varepsilon  = |\psi^\varepsilon|^2 , \\
\label{dm3} &\Box A_k^{\varepsilon}  =  \langle \psi ^\varepsilon, \alpha^k \psi^\varepsilon \rangle,
\qquad k=1, 2, 3,
\end{align}
subject to \emph{Cauchy initial data}:
\begin{equation}
\left \{
\begin{split}
&\psi ^\varepsilon  \big| _{t=0}=\psi ^\varepsilon  _I(x)\sim O(\sqrt{\varepsilon}), \\
&V^\varepsilon \big| _{t=0}=V^\varepsilon_I(x), \quad \partial _{t}V^\varepsilon \big| _{t=0}=
\tilde V^\varepsilon_I (x) ,  \\
&A^{\varepsilon} \big| _{t=0}=A^{\varepsilon}_{I}(x)  ,
\quad \partial _{t}A^{\varepsilon} \big| _{t=0}= \tilde A^{\varepsilon}_{I} (x).
\end{split}
\right.
\end{equation}
For this nonlinear system, we want to find an asymptotic
description of $\psi^\varepsilon\sim O(\sqrt{\varepsilon})$ as
$\varepsilon \rightarrow 0$, \ie a semiclassical description. Note
that, \emph{equivalently}, one could consider asymptotic solutions
of the form
\begin{equation}
\Phi^\varepsilon(t,x):=\frac{1}{\sqrt{\varepsilon}} \, \psi^\varepsilon(t,x)\sim O(1),
\end{equation}
which do not vanish in the limit $\varepsilon \rightarrow 0$ and
which again satisfy the semiclassical scaled DM system, modified
by the fact that the right hand sides of (\ref{dm2}), (\ref{dm3})
are multiplied by an additional factor $\varepsilon$. This
illustrates the fact that we are dealing with a \emph{small
coupling limit}. We further stress that in our scaling the mass is
$O(1)$ and \emph{fixed} as $\varepsilon \rightarrow 0$, which is
different from the otherwise similar scaling used in \cite{KS},
where a classical mechanics analogue of the DM system has been
studied.
\newpar
The (rigorous) analysis of semiclassical asymptotics has a long tradition in quantum mechanics, the most
common technique being the so called \emph{WKB-method}. Quite
recently, a semiclassical approach to the \emph{linear} Dirac
equation was taken in \cite{BK} and also, using \emph{Wigner
measures}, in \cite{GMMP}, \cite{Sp}. For a broader introduction
on linear techniques and results, we refer to \cite{MaFe},
\cite{Ro}, \cite{SMM} and the references given therein. Nonlinear extensions
of the WKB-method can be found for example in \cite {Ge},
\cite{Gr}, where \emph{scalar-valued} semilinear Schr\"odinger
equations are analyzed. We remark that the case of the nonlinear
Dirac $4$-system introduces significant new difficulties in the
WKB-analysis, some of them are already present in the linear
setting.
\newpar
Mathematically, our approach is inspired by techniques developed
in \cite{DoRa}, which sometimes go under the name \emph{weakly
nonlinear geometrical optics}. Due to the appearance of the small
parameter $\varepsilon$ in front of each derivative in (\ref{dm1})
we are in the regime of so-called \emph{dispersive} weakly
nonlinear geometrical optics, which differs in several aspects
from the non-dispersive one. We remark that the latter case is
much better studied in the so far existing literature and we refer
to \cite{JMR2}, for a recent review.
\newpar
To be more precise, we shall seek a local-in-time solution of (\ref{dm1}), which asymptotically takes the following form:
\begin{equation}
\left \{
\begin{split}
\label{ex1} \psi ^\varepsilon (t,x)= & \ \sqrt{\varepsilon}\, u^\varepsilon (t,x, \phi(t,x)/\varepsilon),\\
 u^\varepsilon (t,x, \theta) \sim & \ \sum_{j=0}^\infty  \varepsilon ^{j/2} u_j(t,x,\theta ).
\end{split}
\right.
\end{equation}
Here, all the $u_j(t,x,\theta )\in \mathbb C^4$ being $2\pi $-periodic w.r.t. $\theta \in \mathbb R$.
Due to the factor $\varepsilon ^{1/2}$, we call them \emph{small semiclassical approximate solutions},
or \emph{small WKB-solutions}.
\newpar
As expected, the phase function $\phi $ satisfies a (free)
relativistic Hamilton-Jacobi equation and of course, convergence
of the expansion (\ref{ex1}), can only hold on a time interval,
which corresponds to the existence of smooth solutions $\phi $.
In weakly nonlinear geometrical optics, the homogeneity of the nonlinearity
determines the required order of smallness of the asymptotic
solution and, as we shall see, the factor $\sqrt{\varepsilon}$ precisely fits with
the cubic nonlinearities in (\ref{dm1}).
We will show that for this particular scale we obtain,
on the one hand, \emph{independent} propagation of the electronic
resp. positronic phase function $\pm \phi $ and, the other hand,
\emph{nonlinear interaction} of the corresponding principal
amplitudes $u_{0,\pm}$.
\newpar
In other words, we study solutions on the threshold of
\emph{adiabatic decoupling}, a phenomena which is already well
known in the linear case, \cf  \cite{PST}. In particular, the
importance of the $O(\sqrt{\varepsilon })$-scale for the (linear)
Dirac equation is stressed in \cite{FK}, where one can also find a
detailed description of the energy-transfer between electrons and
positrons in terms of \emph{two-scale} Wigner measures. These
results, together with ours suggest that if one wants to obtain
semiclassical $O(1)$-approximation in the strongly coupled regime,
one needs to take into account simultaneously scales of order
$\varepsilon$ and $O(\sqrt{\varepsilon })$. These asymptotic
solutions are then appropriate for heavily charged particles. We
finally remark, that in the very recent paper \cite{Je}, coupled
Gauge-fields are studied from a similar point of view as in our
work.
\newpar
This paper is organized as follows: We collect some preliminaries
in section 2, then we shall determine the critical exponent and
the corresponding eikonal equation of the approximate WKB-type
solution in section 3. The corresponding
$\varepsilon$-oscillations introduced by the nonlinearity are
determined in section 4 and the nonlinear transport of the
approximation along the rays of geometrical optics is obtained in
section 5. Finally, in section 6 we shall prove that there exists
a (local-in-time) solution of the MD system which stays close to
the approximation and we also collect some further qualitative
results.


\section{Preliminaries}

In the following, we will assume that no external electromagnetic fields are present:
\begin{equation}
V ^{ext} (t,x) \equiv 0, \quad A ^{ext}(t,x)\equiv 0.
\end{equation}
Moreover, we assume that at $t=0$ we have:
\begin{equation}
V ^\varepsilon _I (x) =\tilde V ^\varepsilon _I (x) \equiv 0, \quad A ^\varepsilon _{I}(x)=
\tilde A ^\varepsilon _{I}(x) \equiv 0.
\end{equation}
Neither of these assumptions changes the following analysis
significantly. They are only imposed for the sake of simplicity.
Note that the DM system is time-reversible, but w.r.o.g. we shall
consider positive times only in the sequel.
\newpar
Using the fundamental solution of the wave equation in dimension $d=3$ and for times $t>0$,
we find the following expression for $V$, called \emph{the retarded potential}:
\begin{align}
\label{wso1}
V^\varepsilon[\psi ^\varepsilon  ] (t,x)
& = \ \frac{1}{4\pi} \int_{|x-y|\leq t }\frac{|\psi^\varepsilon (t-|x-y|,y)|^2 }{|x-y|} \ dy \\
& =: \ \mathcal G _r (t,x) \ast |\psi^\varepsilon (t,x)|^2 ,\nonumber
\end{align}
where $\ast $ denotes the convolution w.r.t. $(t,x)$ and
\begin{equation}
\mathcal G_r (t,x) :=\frac{\Theta (t)}{4\pi |x|}\ \delta (t-|x|).
\end{equation}
Likewise, $A$ can be written as:
\begin{align}
\label{wso2} A^\varepsilon[\psi ^\varepsilon  ] (t,x) =   & \ \mathcal G _r  (t,x) \ast
\langle \psi ^\varepsilon (t,x), \alpha^k \psi^\varepsilon(t,x)\rangle.
\end{align}
Using this representations, we shall rewrite (\ref{dm1})-(\ref{dm3}) in the form of a \emph{semilinear Dirac equation}:
\begin{equation}
\label{dm}
\left \{
\begin{split}
i\varepsilon \partial _t\psi ^\varepsilon - \mathcal D^\varepsilon _A (t,x,\varepsilon D_x)\psi ^\varepsilon =
&\ 0,\qquad x\in \mathbb R^3, \ t>0,\\
\psi ^\varepsilon \big| _{t=0}=& \ \psi^\varepsilon  _I(x),
\end{split}
\right.
\end{equation}
where $\mathcal D^\varepsilon _A$ is a \emph{matrix-valued} differential operator ($D_x:=-i \nabla $).
The corresponding $\varepsilon$\emph{-dependent} symbol is given by
\begin{align}
\mathcal D^\varepsilon _A(t, x,\xi ) = \alpha\cdot (\xi - A^\varepsilon[\psi ^\varepsilon](t,x))+\beta +
V^\varepsilon[\psi ^\varepsilon ](t,x),
\quad
\end{align}
where $x,\xi,\in \mathbb R^{3}, t\in \mathbb R$. Here and in the following we use the notation
\begin{align}
\alpha  \cdot \xi  := \sum_{k=1}^3 \alpha ^k \xi _k.
\end{align}
Note that in the nonlinear equation (\ref{dm}),
the potentials $A^\varepsilon [\psi ^\varepsilon ]$, $V^\varepsilon [\psi ^\varepsilon ]$ depend
\emph{non-locally} on $\psi ^\varepsilon $,
as indicated by the bracket-notation.
\newpar
Multiplying (\ref{dm}) with $\psi ^\varepsilon $ and taking imaginary parts,
we obtain the usual conservation law for ${\| \psi ^\varepsilon (t,x) \|}_2$, hence the \emph{conservation of charge}:
\begin{equation}
\int_{\mathbb R^3} \langle \psi ^\varepsilon (t,x), \psi ^\varepsilon (t,x)\rangle \ dx =
{\| \psi ^\varepsilon (t,x) \|}_2^2 =  \mbox {const.}
\end{equation}
The \emph{free Dirac operator} will be denoted by
\begin{align}
\label{do}\mathcal D(\varepsilon D_x)\psi ^\varepsilon := -i\varepsilon (\alpha \cdot \nabla)
\psi ^\varepsilon +\beta \psi ^\varepsilon,
\end{align}
with symbol
\begin{align}
\label{sym}\mathcal D(\xi ) = \alpha \cdot  \xi  +\beta .
\end{align}
This $4\times 4$ matrix has two different eigenvalues $h_\pm(\xi) $ of multiplicity $2$ each:
\begin{align}
\label{ham}h_\pm(\xi ):= \pm \lambda (\xi ) ,\quad \xi \in\mathbb R^3,
\end{align}
where
\begin{align}
\label{ev} \lambda (\xi ) :=\sqrt{{|\xi|}^2+1},\quad \xi \in\mathbb R^3.
\end{align}
As expected, the eigenvalues $h_\pm (\xi) $ are nothing but the \emph{free Hamiltonian} for a \emph{relativistic}
particle.
The positive resp. negative sign in (\ref{ev}) corresponds to \emph{electrons} resp. \emph{positrons}.
By straightforward calculations we obtain:
\begin{lemma}
The \emph{spectral projectors} $\Pi_\pm(\xi  ):\mathbb C^4\rightarrow \mathbb C^4$, associated to $h_\pm(\xi) $
are given by
\begin{align}
\label{pro}\Pi_\pm  (\xi ):=\frac{1}{2}\left(\id_4\pm \frac{ 1}{\lambda(\xi )}\
\mathcal D (\xi )\right),\qquad \Pi_\pm \Pi_\mp \equiv 0.
\end{align}
\end{lemma}
The matrix-valued symbol $\mathcal D(\xi)$ can therefore be decomposed into its positive
and negative energy part in the following way:
\begin{align}
\label{dec}
\mathcal D(\xi ) = h_+ (\xi) \Pi_+ (\xi ) +h_-(\xi)\Pi_-(\xi ) .
\end{align}
Note that
\begin{equation}
\label{rel1}h_\pm(-\xi )=h_\pm(\xi ),\quad\mbox{whereas \ } \Pi_\pm(-\xi )=\Pi_\mp(\xi ).
\end{equation}
For later purpose, we also define:
\begin{definition}
The \emph{partial inverse} $ \Lambda _\pm( \xi ):\mathbb C^4\rightarrow \mathbb C^4$,
associated to $\Pi_\pm(\xi  )$ is given by
\begin{align}
\label{pain}\Lambda _\pm( \xi )\Pi_\pm( \xi ) =0, \qquad \Lambda _\pm( \xi )\mathcal D( \xi )X =(\id_4 -\Pi_\pm(\xi ))X ,
\quad \forall X \in \mathbb C^4.
\end{align}
\end{definition}
Finally, we recall the definition of \emph{asymptotic equivalence}:
\begin{definition}
Let $\mathcal O\subseteq \mathbb R^n$, $n\geq1$, be an open set, $a_j(y)\in C^\infty (\mathbb R^n; \mathbb C^n)$ and
$a^\varepsilon \in C^\infty (]0,\varepsilon _0[\times \mathbb R^n; \mathbb C^n)$. Then we say that $a^\varepsilon $ is
asymptotically equivalent to the formal sum $\sum_{j=0}^\infty \epsilon ^j a_j$ and write
\begin{equation}
a^\varepsilon (y)\sim \sum_{j=0}^\infty \varepsilon ^j a_j(y),
\end{equation}
if for every $m>0$, every multiindex $\sigma$ and every compact subset $\mathcal K\subset \mathcal O$,
there exists a $C_{m,\sigma}>0$, such that
\begin{equation}
\sup_{\mathcal K} \left| \partial ^\sigma _y \left( a^\varepsilon (y)-\sum_{j=0}^m \varepsilon ^j a_j(y)\right)\right|
\leq C_{m,\sigma} \varepsilon ^m.
\end{equation}
\end{definition}


\section{Generalized WKB-Ansatz and the Eikonal Equation}

At first we will show that the desired $O(\sqrt {\varepsilon})$-asymptotics for the spinor field fits into
the framework of
weakly nonlinear (dispersive) geometrical optics, as introduced in \cite{DoRa}, for nonlinear hyperbolic systems.
\newpar
We plug the following \emph{generalized WKB - Ansatz} into equation (\ref{dm}):
\begin{equation}\label{an}
\left \{
\begin{split}
\psi ^\varepsilon (t,x)= & \  \varepsilon ^p \, u^\varepsilon (t,x, \phi(t,x)/\varepsilon),\\
 u^\varepsilon (t,x, \theta) \sim & \ \sum_{j=0}^\infty  \varepsilon ^{jp} u_j(t,x,\theta ),
\end{split}
\right.
\end{equation}
where the functions $u_j(t,x,\theta )\in \mathbb C^4$ are assumed to be sufficiently smooth and
$2\pi $-periodic w.r.t. $\theta \in \mathbb R$.
This gives
\begin{align}
\label{f}
0= & \ i\varepsilon^{p+1} \partial _t u^\varepsilon -  \varepsilon ^p \,
\mathcal D^\varepsilon _A (t,x,\varepsilon D_x) u ^\varepsilon \\
 = & \  i \varepsilon^{p+1} (\partial _t u^\varepsilon  + (\alpha \cdot \nabla) u^\varepsilon )  -
 \varepsilon ^p \beta u^\varepsilon +
i  \varepsilon ^p \, (\partial _t \phi  + (\alpha \cdot \nabla \phi) ) \partial _\theta u^\varepsilon +
\varepsilon ^{3p}\mathcal N^\varepsilon [u^\varepsilon ],\nonumber
\end{align}
with a \emph{nonlinearity}, $\mathcal N^\varepsilon :\mathbb C^4\rightarrow \mathbb C^4$, defined by
\begin{equation}
\label{nl} \mathcal N^\varepsilon[u^\varepsilon ]:=  \left((\alpha \cdot A^\varepsilon [u^\varepsilon ])-
V^\varepsilon[u^\varepsilon ]\right)u^\varepsilon .
\end{equation}
The strategy is now to expand the right hand side of (\ref{f}) as
\begin{equation}\label{f1}
\varepsilon^p \sum_{j=0}^\infty  \varepsilon ^{jp} R_j(t,x)
\end{equation}
and choose the coefficients $u_j$ of (\ref{an}) in such a way, that $R_j (t,x)\equiv 0$, $\forall j\in \mathbb N $.
\newpar
It is important to note that the first term on the right hand side
of (\ref{f}) is of order $\varepsilon ^{p+1}$, whereas the second
and the third are $\sim O(\varepsilon ^p)$. Since $A^\varepsilon
[u^\varepsilon ]$, $V^\varepsilon [u^\varepsilon ]$ are of order
$\varepsilon ^{2p}$, by equation (\ref{wso1}), (\ref{wso2}), the
function $\mathcal N^\varepsilon  [u^\varepsilon ]$ is of order
$\varepsilon ^{3p}$. This nonlinear term is supposed to be
\emph{small}, more precisely, it should not enter into the
equation for $R_0$, describing  terms of order $O(\varepsilon
^p)$, but rather into expressions of $O(\varepsilon ^{p+1})$.
Thus we are led to the following \emph{normalization condition}:
\begin{equation}
3p = p+1,
\end{equation}
implying $p=1/2$. With this normalization we have $u^\varepsilon \sim O(\varepsilon ^{1/2})$
(just as required by the scaling
presented in the introduction), whereas the nonlinear term satisfies:
$\mathcal N^\varepsilon [u^\varepsilon ]\sim O(\varepsilon ^{3/2})$.
\begin{remark} The choice $p=1/2$  gives the critical exponent in the sense that for amplitudes $O(\varepsilon ^{1/2})$
one can prove simultaneously existence of the approximate smooth solution for times $t=O(1)$,
\ie on a time-scale \emph{independent} of $\varepsilon $,
and nontrivial nonlinear behavior in the principal term of the approximation, \cf \cite{DoRa}, \cite{JMR2}.
\end{remark}
Setting $R_0(t,x)=0$, yields
\begin{equation}
i( \partial _t \phi  +  (\alpha \cdot \nabla \phi )) \partial _\theta u_0 - \beta u_0 =0.
\end{equation}
Since $u_j\in C^\infty(\mathbb R^4\times \mathbb S^1;\mathbb C^4)$ it can be \emph{Fourier-expanded} w.r.t. $\theta $
\begin{equation}
\label{fex}
u_0(t,x,\theta )=\sum_{m\in \mathbb Z} u_{0,m}(t,x)e^{im\theta }.
\end{equation}
By this procedure, we find the following equation for the coefficients $u_{0,m}$:
\begin{equation}
\label{zeq}
L( m \, d\phi(t,x))u_{0,m}:= \left(m \, \partial _t \phi +  \mathcal D (m \nabla \phi)\right) u_{0,m} =0,
\end{equation}
where $\mathcal D (\nabla (m\phi ))$ is the $4\times 4$ symbol matrix of the free Dirac operator evaluated at
$\xi =\nabla (m\phi (t,x))$.
In order to have a nontrivial solution $u_{0,m}\not=0$ we impose the condition, that
there exists an open set $\Omega \subseteq \mathbb R^{1+3}$, having a nontrivial intersection with $\{t=0\}$, s.t.
\begin{equation}
\det L( m\, d \phi(t,x))=0,\quad \forall (t,x)\in \Omega \subseteq \mathbb R^{1+3}.
\end{equation}
Using equations (\ref{ham}), (\ref{dec}), this is equivalent to
\begin{equation}
(m\, \partial _t \phi)^2 - |m\nabla \phi |^2=1,\quad \forall (t,x)\in \Omega \subseteq \mathbb R^{1+3}.
\end{equation}
Thus, for $m=\pm1$, the phase function $\phi$ satisfies (in $\Omega $)
the \emph{eikonal equation} for the \emph{Klein-Gordon operator}, \ie
\begin{equation}
\label{eik} (\partial _t \phi)^2 - |\nabla \phi |^2=1,\quad \forall (t,x)\in \Omega \subseteq \mathbb R^{1+3}.
\end{equation}
Indeed, it is easy to see that the choices $m=\pm1$ are the only possibilities, since equation (\ref{eik}) gives
\begin{align}
 (m\partial _t \phi)^2 - |m\nabla \phi |^2-1= & \ (m^2-1)((\partial _t \phi)^2 - |\nabla \phi |^2) \\
= & \ m^2-1, \nonumber
\end{align}
which is different from zero for all $m\not=\pm1$.
Hence, in the Fourier-series (\ref{fex}), there appear only two nontrivial \emph{harmonics},
which are associated to the eikonal equation (\ref{eik}): namely $\exp(i\phi (t,x)/\varepsilon )$,
for $m=1$ and $\exp(-i\phi(t,x)/\varepsilon ) $, for $m=-1$.\\
For $m=\pm1$ the equation (\ref{eik}) is fulfilled by two possible $\phi $'s, obtained from
\begin{align}
\label{hj}\partial _t \phi_\pm(t,x)= h_\pm(\nabla \phi _\pm(t,x))\equiv \pm\sqrt{{| \nabla \phi_\pm (t,x)|}^2+1}=0.
\end{align}
This is the \emph{Hamilton-Jacobi equation} for free relativistic particles. The following lemma guarantees existence
and uniqueness of smooth solutions, where from now on, we shall denote by $ D^2 f(x) $, the Hessian
of a given function $f:\mathbb R^3 \rightarrow \mathbb R$.
\begin{lemma} \label{le0}
Given $\phi_I \in C^\infty(\mathbb R^3;\mathbb R)$, s.t. $\| D^2\phi_I(x) \|\leq C$,
there exist $T_\pm>0$ and uniquely determined functions
$\phi_\pm\in C^\infty (\Omega_\pm ; \mathbb R)$, where $ \Omega_\pm :=[0,T_\pm)\times \mathbb R^3$, s.t.
\begin{equation}
\label{hj1}
\left \{
\begin{split}
\partial _t \phi_\pm(t,x)=& \ h_\pm(\nabla \phi_\pm (t,x)), \quad \forall (t,x)\in \Omega_\pm,\\
\phi \big |_{t=0}=   & \ \phi_I(x).
\end{split}
\right.
\end{equation}
\end{lemma}
\begin{proof} We only proof the assertion for $\phi _+$, since the other case is completely analogous.
The initial value problem is \emph{non-characteristic} everywhere, since
\begin{equation}
\partial _t\phi_+(0,x) = \sqrt{{| \nabla \phi_+ (0,x)|}^2+1} \ \not=0, \ \forall x\in\mathbb R^3.
\end{equation}
Thus, $\partial _t\phi_+ (0,x)$ can be obtained from the initial data $\phi_+ (0,x)=\phi_I(x)\in C_b^\infty (\mathbb R^3)$
at each point $x\in\mathbb R^3$. Standard PDE theory then guarantees the existence
of a unique smooth solution $\phi_+ \in C^\infty (\Omega_+;\mathbb R) $, as long as
\begin{equation}
1-t \, \| D^2h_+(\nabla \phi_I )\| \, \| D^2\phi_I \|\not=0.
\end{equation}
Since $D^2H(\xi)$ is uniformly bounded, this condition holds by assumption and the assertion is proved.
\end{proof}
\begin{remark} The assumption in lemma \ref{le0} can be relaxed to $\phi_I \in C^\infty(\mathbb R^3;\mathbb R)$.
In this case, however, one can not guarantee the existence a smooth solution $\phi$ in a space-time slab,
but only in some
open set $\mathcal O \subset \mathbb R^{1+3}$. In the following, this would lead to some technical difficulties,
which we want to avoid, though, the whole procedure can be generalized to that case.
\end{remark}
By equation (\ref{hj}), we have
\begin{equation}
\label{rel2}-\phi _\pm(t,x)=\phi _\mp(t,x),
\end{equation}
assuming that it holds initially at $\{t=0\}\subset \Omega$.
In the following, we therefore consider only the solution to (\ref{hj}) with positive sign in front of the square root
and write for it $\phi (t,x)\equiv \phi _+(t,x)$. Also, we henceforth denote by $\Omega :=[0,T_+)\times \mathbb R^3$
the slab, in which existence of a smooth function $\phi$ is guaranteed.
This we can do w.r.o.g. as will become clear in a moment:
\newpar
Since for $\phi\equiv \phi _+$ it holds that $\partial _t \phi- h_+(\nabla \phi)=0$, equation (\ref{zeq}) implies the
following \emph{polarization conditions}, locally for all $(t,x)\in \Omega$:
\begin{equation}
\label{po1}(\Pi_-(\nabla \phi)u_{0,+1})(t,x)=0 \ \Leftrightarrow \ (\Pi_+(\nabla \phi )u_{0,+1})(t,x)=u_{0,+1}(t,x).
\end{equation}
Likewise, we get
\begin{equation}
\label{po2}(\Pi_+(\nabla \phi )u_{0,-1})(t,x)=0 \ \Leftrightarrow \ (\Pi_-(\nabla \phi )u_{0,-1})(t,x)=u_{0,-1}(t,x).
\end{equation}
One easily checks, using (\ref{rel1}) and (\ref{rel2}), that the conditions obtained with the choice $\phi =\phi _-$,
are \emph{equivalent} to (\ref{po1}), (\ref{po2}).
Thus, equation (\ref{zeq}) indeed carries two degrees of freedom for the phase, given by $\pm \phi $
(or equivalently $\phi _+$, $\phi _-$). The amplitudes are then rigidly linked, by (\ref{po1}), (\ref{po2}).
\newpar
In summary, we find that the \emph{principal term} $u_0(t,x,\theta )$ in our asymptotic description is given by
\begin{align}
\label{u0}  u_0(t,x,\phi (t,x)/\varepsilon  ):=  u_{0,+1}(t,x)e^{i \phi (t,x)/\varepsilon }+u_{0,-1}(t,x)e^{-i \phi (t,x)/\varepsilon},
\end{align}
where the amplitudes are polarized according to (\ref{po1}), (\ref{po2}). From  now on we shall use the simplified notation
$u_{0,\pm1}=u_{0,\pm}$.


\section{Oscillations of the Nonlinearity}

Let us determine the response of the wave equations (\ref{dm2}), (\ref{dm3}) to
r.h.s. source terms induced by functions of the form (\ref{u0}):\\
To this end, we calculate:
\begin{align}\label{prho}
|u_0(t,x,\phi(t,x)/\varepsilon)|^2 = |u_{0,+}(t,x)|^2+|u_{0,-}(t,x)|^2.
\end{align}
The terms, which mix the electronic and positronic components cancel, since $\Pi _\pm $ is hermitian
and $\Pi _\pm \Pi _\mp \equiv 0$. Hence, we get  from (\ref{wso1}) (at least formally),
that the scalar potential $V$ generated by the principal term $u_0$, is simply given by
\begin{align}\label{V0}
V[u_0]=\  \mathcal G_r (t,x) \ast  \left(|u_{0,+}(t,x)|^2+|u_{0,-}(t,x)|^2\right).
\end{align}
In order to calculate the magnetic potential corresponding to $u_0$, we first note that,
by definition, we have the following identity
\begin{equation}\label{m1}
\Pi  _\pm(\xi) \left( \alpha \cdot \xi +\beta \right) =h_\pm (\xi )\ \Pi _\pm(\xi)  .
\end{equation}
Differentiating w.r.t. $\xi_k$ and multiplying (from the right) with $\Pi_\pm(\xi)$ gives
\begin{align}\label{m2}
\Pi _\pm (\xi) \alpha^k \Pi _\pm(\xi) = & \
\Pi _\pm (\xi) (\partial _{\xi _k} h_\pm (\xi ))\Pi _\pm(\xi) \\
 = & \pm \frac{\xi_k}{\sqrt{{|\xi |}^2+1}} \ \Pi _\pm(\xi) ,\nonumber
\end{align}
since $\Pi^2_\pm(\xi)=\Pi_\pm(\xi)$. The expression
\begin{equation}
\label{vel} \omega _\pm(\xi):=\nabla _\xi  h_\pm (\xi )= \pm \frac{\xi}{\lambda(\xi)},
\end{equation}
is called the electronic resp. positronic \emph{group velocity}, $\omega_\pm \in C^\infty(\mathbb R^3;\mathbb R^3)$.
Using this definition, we obtain for $k=1,2,3$:
\begin{align}\label{j0}
& \langle u_0(t,x,\phi (t,x)/\varepsilon  ), \alpha^k u_0(t,x,\phi (t,x)/\varepsilon  )\rangle  \\
& = \ \omega _{+,k}(\nabla \phi(t,x) )|u_{0,+}(t,x)|^2  \ + \omega _{-,k}(\nabla \phi(t,x)) |u_{0,-}(t,x)|^2  \nonumber \\
&  \ + \langle{u}_{0,+}(t,x), \alpha^k u_{0,-}(t,x)\rangle e^{-i2\phi (t,x)/ \varepsilon}  \ \nonumber \\
&  \ + \langle{u}_{0,-}(t,x), \alpha^k u_{0,+}(t,x)\rangle e^{i2\phi(t,x) / \varepsilon}.\nonumber
\end{align}
The oscillating terms are usually called the \emph{Zitterbewegung} of the Dirac-current, \cf \cite{Sc}, p.$\,195$.
The fact that the current-density corresponding to $u_0$ carries $\varepsilon$-oscillations
is in sharp contrast to the WKB-approach for Schr\"odinger-type problems, see \eg \ \cite{Ge}, \cite{Gr}.
\newpar
The Zitterbewgung may cause severe problems since \emph{a-priori}
one cannot exclude the possibility of \emph{resonant interactions}
between the principal term $u_0$ and the magnetic potential
$A^\varepsilon [u_0]$ obtained from (\ref{dm3}) with r.h.s. given
by (\ref{j0}). If this happens to be the case, our one-phase
ansatz (\ref{an}) breaks down and instead one would need to
establish a so-called \emph{resonant asymptotic expansion} in the
spirit of \cite{JMR1}. (We remark that so far, only the case of
resonances in one spatial dimension can be treated rigorously, \cf
\cite{JMR2}.)
\newpar
We will show that these problems do not appear in our situation.
To this end, we need to describe precisely what kind of
$\varepsilon$-oscillations are present in $A^\varepsilon [u_0]$.
\newpar
First we note that, by the superposition principle, every term appearing on the r.h.s. of (\ref{j0}) generates its own potential field.
The nonoscillating terms of (\ref{j0}) lead to a standard hyperbolic problem, hence (\ref{wso2}) gives
\begin{align} \label{A0}
A_{0}[u_0] (t,x):= \mathcal G_r(t,x) \ast  \left ( \omega_{+}(\nabla \phi )|u_{0,+}|^2+ \omega_{-,k}(\nabla \phi) |u_{0,-}|^2 \right)(t,x).
\end{align}
In order to treat the Zitterbewegung, let us define
$Z := ( Z_1, Z_2, Z_3)$ by
\begin{align}\label{z}
Z_k (t,x,\phi(t,x)/\varepsilon):= \ & \langle{u}_{0,+}(t,x), \alpha^k u_{0,-}(t,x)\rangle e^{-i2\phi (t,x)/ \varepsilon} \\
\ & + \langle{u}_{0,-}(t,x), \alpha^k u_{0,+}(t,x)\rangle e^{i2\phi(t,x) / \varepsilon},\quad k=1,2,3. \nonumber
\end{align}
Using this definition, we can now prove the following lemma:
\begin{lemma} \label{le1}
Let $\Omega \subseteq \mathbb R^{1+3}$ be the slab in which existence of a smooth phase
$\phi \in C^\infty (\Omega ;\mathbb R)$, satisfying (\ref{hj1}), is guaranteed.
Then, given $Z\in C^\infty (\Omega \times \mathbb S^1 ; \mathbb C^3)$, as in (\ref{z}),
there exists a uniquely determined smooth $A^\varepsilon\in C^\infty (\Omega \times \mathbb S^1; \mathbb C^3)$, with
\begin{align}\label{aasy}
A^\varepsilon (t,x,\theta  ) \sim \sum_{l =1}^\infty  \varepsilon ^l A_{l} (t,x,\theta ),
\end{align}
s.t. $A^\varepsilon (t,x,\phi(t,x)/\varepsilon)$ satisfies:
\begin{equation}
\left \{
\begin{split}
& \Box A^\varepsilon(t,x) - Z (t,x) \sim 0, \quad \mbox{ in $C^\infty (\Omega;\mathbb C^3)$},\\
& A^\varepsilon   \big |_{t=0}=  \partial _t A^\varepsilon  \big |_{t=0}=  \ 0.
\end{split}
\right.
\end{equation}
More precisely, $A_i(t,x, \phi (t,x)/\varepsilon )$ can be written in the following form:
\begin{align}\label{A1}
A_l (t,x, \phi (t,x)/\varepsilon ) = A_l^+(t,x)e^{ i2  \phi (t,x)/\varepsilon } + A_l^- (t,x)e^{-i2 \phi (t,x)/\varepsilon },
\end{align}
with principal amplitudes $A_1^\pm\in C^\infty (\Omega; \mathbb C^3)$, given by
\begin{align}\label{apr}
A^\pm_{1,k}(t,x)= \pm \ \langle u_0(t,x), \alpha^k u_0(t,x)\rangle.
\end{align}
\end{lemma}
\begin{proof} The proof can be done separately for each spatial component of $A^\varepsilon (t,x,\theta)$ and for
both types of oscillations, corresponding to $\pm 2\phi$. Hence, we are lead to the following type of problem:
\begin{equation}\label{ap1}
\left \{
\begin{split}
& \Box a^\varepsilon (t,x) =  b(t,x) e^{\pm i2 \phi (t,x)/\varepsilon }, \quad \mbox{ $(t,x) \in \Omega ,$}\\
& a^\varepsilon  \big |_{t=0}=  \partial _t a^\varepsilon  \big |_{t=0}=  \ 0,
\end{split}
\right.
\end{equation}
for some given $b \in C^\infty (\Omega ; \mathbb C)$. Let us define a new variable $f^\varepsilon (t,x)\in \mathbb C^5$ by
$$
f^\varepsilon (t,x) :=( \partial _1 a^\varepsilon ,\partial _2 a^\varepsilon ,\partial _3
a ^\varepsilon ,\partial _t a ^\varepsilon , a^\varepsilon )^\top (t,x).
$$
Further, denoting by
$$
\hat b(t,x):= (0,0,0,b(t,x),0)^\top ,
$$
we can rewrite (\ref{ap1}) in the form of a symmetric hyperbolic system
\begin{equation}\label{ap2}
\left(\partial _t + (\lambda \cdot  \nabla) +\kappa \right) f^\varepsilon (t,x) = \hat b(t,x)e^{\pm i2 \phi (t,x)/\varepsilon },
\end{equation}
with $\lambda  ^k$, $\kappa$, denoting realvalued (symmetric) $5\times 5$ matrices.
In our case, these matrices are simply given by (see \eg \ \cite{R}, p.$\,21$, for more details):
\begin{equation}
\lambda ^k:=-\left( \delta _{mk}\delta _{n4}+\delta _{m4}\delta _{nk}\right)_{m,n},
\quad \kappa :=-\left( \delta _{5m}\delta _{4n}\right)_{m,n}, \quad m,n=1,\dots,5,
\end{equation}
where $\delta _{ab}$ denotes the Kronecker symbol and $k=1,2,3$.
\newpar
It is now possible to use the existing results on linear
geometrical optics, provided the phase $\pm 2\phi $ is \emph{not}
characteristic for the system (\ref{ap2}), \ie
\begin{equation}\label{ap3}
\det \left(\pm 2 \partial _t \phi (t,x) \pm 2\lambda \cdot \nabla  \phi (t,x)  +
\kappa \right)\not=0, \quad \forall (t,x)\in \Omega .
\end{equation}
Computing this determinant, we obtain the condition
\begin{equation}\label{ap4}
\pm 32\, (\partial _t \phi)^3 ((\partial _t \phi)^2 - |\nabla \phi |^2)\not=0,\quad \forall (t,x)\in \Omega.
\end{equation}
Since, by assumption, $\phi $ solves the Klein-Gordon eikonal equation (\ref{eik}) the second factor on the l.h.s. of (\ref{ap4})
is equal to one and thus, different from zero in all of $\Omega $.
On the other hand we get from (\ref{eik}): $(\partial _t \phi)^3 = (|\nabla \phi |^2+1)^{3/2}\not= 0$,
$\forall (t,x)\in \Omega$. Hence, condition (\ref{ap3}) is fulfilled and the assertion follows from theorem 4.4 in \cite{Ra}.
In particular we get:
\begin{align}
A^\pm_{1,k}(t,x)= & \ \frac{\mp 1}{-(\partial _t \phi)^2 + |\nabla \phi |^2}\  \langle u_0(t,x), \alpha^k u_0(t,x)\rangle \\
= & \ \pm \ \langle u_0(t,x), \alpha^k u_0(t,x)\rangle \nonumber,
\end{align}
which concludes the proof.
\end{proof}
Lemma \ref{le1} shows that the Zitterbewegung in (\ref{j0})
generates a magnetic potential which is small, \ie at least of
order $O(\varepsilon)$. Moreover the $ßvarepsilon$-oscillations,
appearing in $A^\varepsilon$, are exactly the \emph{same} as in
the (\ref{j0}) and hence we can consistently proceed with our
one-phase expansion for $\psi ^\varepsilon $.
\begin{remark} Although the MD system is hyperbolic, lemma \ref{le1} can be considered as an
analogue of so-called \emph{elliptic} high frequency asymptotics,
the main feature of which is the fact that asymptotic solutions
can be obtained by \emph{local} (in $t, x$) algebraic relations.
In other words, the Maxwell system can be considered
\emph{transparent} w.r.t to the oscillations generated by the
Dirac equation.
\end{remark}
The result of lemma \ref{le1} implies that the nonlinearity $\mathcal N^\varepsilon[u_0]$,
defined in (\ref{nl}), admits an asymptotic expansion of the form
\begin{align}
\label{n0ex}
\mathcal N^\varepsilon[u_0] (t,x,\theta  ) \sim   \mathcal N_0[u_0](t,x,\theta )+
 \sum_{l=1}^\infty \varepsilon ^l \mathcal N_l(t,x,\theta ),
\end{align}
where, using the expressions (\ref{V0}), (\ref{A0}), we have:
\begin{align}\label{N0}
\mathcal N_0(t,x,\theta )= \ \left((\alpha \cdot A_0[u_0](t,x))- V[u_0](t,x)\right)  \, u_0(t,x,\theta )
\end{align}
and for all $l \geq 1$:
\begin{align}
\mathcal N_l(t,x,\theta ):=  \ (\alpha \cdot A_{l}(t,x,\theta)) \, u_0(t,x,\theta ),
\end{align}
with $A_l$ given by (\ref{A1}).
\newpar
Note that the expression (\ref{n0ex}) represents two kinds of $\varepsilon$-oscillations:
Those described by phase-functions $ \pm \phi$ are present in all terms $\mathcal N_l$ with $l\geq 0$,
whereas $\varepsilon$-oscillations with phases $\pm 3 \phi$ appear in $\mathcal N_l$ with $l>0$.
Also, note that $u_0$ enters in a \emph{nonlocal} way only in the lowest order term (\ref{N0}).


\section{Nonlinear Transport along Rays}

We need to find an evolution equation, which determines $u_{0}$ from the initial data.
To this end, let us define an operator $\mathbb P$, which projects on the set of harmonics
corresponding to solutions of the eikonal equation (\ref{eik}):
\begin{definition} \label{defp} Given some $v\in C^\infty(\mathbb R^4\times \mathbb S^1;\mathbb C^4)$,
which can be represented by
\begin{equation}
v(t,x,\theta )=\sum_{m\in \mathbb Z} v_{m}(t,x)e^{im\theta },
\end{equation}
we define the action of $\mathbb P$ on $v$, by
\begin{align}
( \mathbb Pv)(t,x,\theta ):=  (\Pi_+(\nabla \phi )v_{+1})(t,x)e^{i\theta }+(\Pi_-(\nabla \phi )v_{-1})(t,x)e^{-i\theta }.
\end{align}
\end{definition}
In words: $ \mathbb P$ picks modes corresponding to $m=\pm1$ and multiplies them with the matrices $\Pi _{\pm}(\nabla \phi )$.
Note that, at least in $\Omega $, it holds true that
\begin{equation}\label{poid}
( \mathbb P  u_0)(t,x,\theta ) =  u_0(t,x,\theta ),
\end{equation}
in view of (\ref{zeq}) and (\ref{po1}), (\ref{po2}).
\newpar
From (\ref{f}), we have that the evolution of $u_0$ is determined by terms of order $O(\varepsilon^{p+1})=O(\varepsilon^{3/2})$.
Setting the corresponding coefficient in (\ref{f1}) equal to zero, \ie $R_2(t,x)  =0 $, yields
\begin{align}\label{noeq}
i( \partial _t \phi  +  (\alpha \cdot \nabla \phi )) \partial _\theta u_2-\beta u_2+
i (\partial _t u_0 +(\alpha \cdot \nabla ) u_0) +\mathcal N_0[u_0 ] =0,
\end{align}
with $\mathcal N_0[u_0]$ as in (\ref{N0}). Equation (\ref{noeq}) implies
\begin{align}
i (\partial _t+(\alpha \cdot \nabla ))u_0 +\mathcal N_0[u_0 ]\in \ran
\Big( i( \partial _t \phi  +  (\alpha \cdot \nabla \phi )) \partial _\theta - \beta \Big).
\end{align}
Applying $ \mathbb P $ to (\ref{noeq}), eliminates the term including $u_2$, since $ \mathbb P $ projects on the kernel of
$ i( \partial _t \phi  +  (\alpha \cdot \nabla \phi )) \partial _\theta - \beta $ and we obtain
\begin{align}
i \mathbb P \partial _t u_0+ i \mathbb P  (\alpha \cdot \nabla) u_0 + \mathbb P \mathcal N_0[u_0 ]=0.
\end{align}
Using the fact that $\mathbb Pu_0=u_0$, by (\ref{poid}), this gives
\begin{align}
\label{teq1} \mathbb P \partial _t ( \mathbb P u_0)+  \mathbb P  (\alpha \cdot \nabla )(\mathbb P u_0) =
i \mathbb P \mathcal N_0[ \mathbb P u_0 ].
\end{align}
This equation is similar to the one appearing in \cite{DoRa}, however, in contrast to the quoted work,
our nonlinearity constitutes only the first term of an asymptotic expansion of the full $\mathcal N^\varepsilon[u_0]$.
\newpar
We proceed by stating a useful identity:
\begin{align}
\label{m3} \alpha^k\Pi _\pm (\xi)= \Pi _\mp(\xi) \alpha^k + \omega_{\pm,k}(\xi) \Id{4},
\end{align}
obtained from straightforward calculations. After more lengthy but straightforward calculations, in which we apply the relations
(\ref{m1}), (\ref{m2}) and (\ref{m3}), we can express the l.h.s. of (\ref{teq1}) in the form of a transport operator:
\begin{align}
&\Pi_\pm\partial _t (\Pi_{\pm}u_0)+ \Pi_{\pm} \alpha \cdot \nabla (\Pi_{\pm}u_0) = \\
\label{teq2} &\partial _t  u_{0,\pm} + (\omega_\pm (\nabla \phi ) \cdot \nabla )u_{0,\pm}+
\frac{1}{2}\diverg (\omega_{\pm}(\nabla \phi )) u_{0,\pm}.\nonumber
\end{align}
On the other hand, computing the action of the projector $ \mathbb P $ on the nonlinear term $\mathcal N_0[u_0]$, we get
\begin{align}
\mathbb P \mathcal N_0[ \mathbb P u_0]= &  \ e^{i\phi/\varepsilon} \left( (A_0[u_0]\cdot \omega_{+} (\nabla \phi)) - V[u_0]
\right)\, u_{0,+}\, \\
& \ + e^{-i\phi/\varepsilon} \left((A_0[u_0] \cdot \omega_{-}(\nabla \phi))- V[u_0] \right)\, u_{0,-}\, .\nonumber
\end{align}
Here we have again used (\ref{m2}).
Thus, we finally conclude, that the time-evolution of the \emph{principal amplitudes} $u_{0,\pm}$ is governed by
the following semilinear first-order system:
\begin{equation}\label{trans}
\left \{
\begin{split}
\left( \partial _t  + (\omega_+ (\nabla \phi ) \cdot \nabla) \right) u_{0,+}(t,x)
= \ \Gamma _+[u_0](t,x) \,u_{0,+}(t,x),\\
\left( \partial _t  + (\omega_- (\nabla \phi ) \cdot \nabla) \right) u_{0,-}(t,x)
= \ \Gamma _-[u_0](t,x)\, u_{0,-}(t,x),
\end{split}
\right.
\end{equation}
where
\begin{equation}
\Gamma _\pm[u_0](t,x) := i A_0[u_0] \cdot  \omega_{\pm} (\nabla \phi) -i
V[u_0]-\frac{1}{2}\diverg (\omega_{\pm}(\nabla \phi )).
\end{equation}
By construction, the polarization of $u_{0,\pm}$ is conserved during the evolution. The system $(\ref{trans})$
determines $( \mathbb P u_0)(t,x,\theta)$,
from its initial data $( \mathbb P u_0)(0,x,\theta)$ and since $( \mathbb P u_0)=u_0$,
we have completely constructed $u_0$.
\newpar
Multiplying (\ref{trans}) by $\overline u_{0,+}$ resp. $\overline u_{0,-}$ and integrating by parts, we obtain the important
property of charge-conservation:
\begin{equation}\label{cons}
\int_{\mathbb R^3} |u_{0,+}(t,x)|^2 +|u_{0,-}(t,x)|^2 \ dx = {\| u_0 (t,x) \|}_2^2 =  \mbox {const.}
\end{equation}
Given $u_0$, determined by (\ref{trans}), it remains to construct the higher order terms $u_j(t,x,\theta)$, $j\geq 1$
of our approximate solution.
This can be done by a similar construction as given in \cite{DoRa}:
\newpar
We expand the cubic nonlinearity $\mathcal N^\varepsilon[u^\varepsilon]$ in powers of $\varepsilon $:
\begin{align}\label{cuex}
\mathcal N^\varepsilon[u_0+\varepsilon u_1+ \cdots] \sim  \ \mathcal N^\varepsilon[u_0] +
\varepsilon \mathcal M^\varepsilon [u_0,u_1]  +\cdots,
\end{align}
where, using the definitions (\ref{V0}), (\ref{z}), we easily compute:
\begin{align}\label{n}
\mathcal M^\varepsilon [u_0,u_1]= & \ 2 \, (\mathcal G_r\ast \langle u_0, u_1\rangle) u_0 + V[u_0] u_1
+ \alpha \cdot (\mathcal G_r\ast Z )\, u_1 \\
& \ + \left(\sum_{k=1}^3 \alpha^k \left(\mathcal G_r\ast \langle u_0, \alpha^ k u_1\rangle + \mathcal G_r\ast
\langle u_1, \alpha^ k u_0\rangle\right)\right)  u_0. \nonumber
\end{align}
We need to apply lemma \ref{le1} to all terms appearing on the r.h.s of (\ref{cuex}), which results in
a similar expansion as given in (\ref{n0ex}). Hence, after rearranging terms in powers of $\varepsilon$, we can write
\begin{align}
\mathcal N^\varepsilon[u_0+\varepsilon u_1+ \cdots] \sim  \ \mathcal N_0[u_0] +
\varepsilon  \mathcal N_1 + \varepsilon  \mathcal M_0[u_0,u_1]
+ \cdots.
\end{align}
Consequently, for $j\geq 1$, the $O(\varepsilon^{j/2+1/2})$-coefficient is given by
\begin{align}
R_{j} (t,x)= &  \ i( \partial _t \phi  +(\alpha \cdot \nabla \phi )) \partial _\theta u_j-\beta u_j  +
i( \partial _t   +(\alpha \cdot \nabla )) u_{j-2}
 \\ & \ + \mathcal M_0[u_0, u_{j-2}] + S(u_0,\dots, u_{n<j-2}) ,\nonumber
\end{align}
where, as usual, we impose: $u_n(t,x,\theta)=0$, for all $n<0$.
The source term $S$, only depends on lower order coefficients $u_0,\dots, u_{n<j-2}$. It is obtained by applying lemma
\ref{le1} to higher order terms in the expansion (\ref{cuex}), leading to contributions $\mathcal N_l$ and $\mathcal M_l$
with: $l+1=j/2$.
\newpar
We can now decompose
\begin{equation}
u_j(t,x,\theta)= ( \mathbb P u_j)(t,x,\theta) + (\id_4- \mathbb P )u_j (t,x,\theta).
\end{equation}
Note that in contrast to $u_0$, where, in view of (\ref{poid}), it holds
\begin{equation}
(\id_4- \mathbb P)u_0 (t,x,\theta)=0,
\end{equation}
we can not expect all higher order coefficients $u_j$ to be polarized too.
Hence, we need to determine separately $ \mathbb P u_j$ and $(\id_4- \mathbb P )u_j$.
To this end, we introduce the following definition:
\begin{definition}
Again, let $v(t,x,\theta)$ be given as in definition \ref{defp},
then we define a partial inverse $\mathbb Q$, associated to $ \mathbb P$, by
\begin{align}
(\mathbb Q v)(t,x,\theta ):=  (\Lambda _+ (\nabla \phi )v_{+1})(t,x)e^{i\theta }+(\Lambda _-(\nabla \phi )v_{-1})(t,x)e^{-i\theta },
\end{align}
where $\Lambda _\pm$ is the partial inverse to $\Pi _\pm$, defined by (\ref{pain}).
\end{definition}
Assume now that we already know $u_n$, for $n<j$, then $(\id_4- \mathbb P )u_j$ is determined by setting
$( \mathbb QR_j)(t,x,\theta )=0$.
This gives
\begin{align}
\label{npr}
(\id_4-\mathbb P)u_j =  -\, \mathbb Q\Big( i( \partial _t   + (\alpha \cdot \nabla )) u_{j-2}+
\mathcal M_0[u_0,u_{j-2}]+  S(u_0,\dots, u_{n<j-2}) \Big).
\end{align}
On the other hand, setting $(\mathbb PR_{j+2})(t,x,\theta )=0$, we obtain an evolution equation for $\mathbb Pu_{j}$:
\begin{align}\label{prp}
i \mathbb P\partial _t (\mathbb Pu_{j})+ i \mathbb P(\alpha \cdot \nabla \mathbb Pu_{j}) =
- \mathbb P \mathcal M_0[u_0,u_{j} ]+ r (u_0, \dots,u_j),
\end{align}
where
\begin{align}
r(u_0,\dots,u_j):= -\mathbb P   S(u_0,\dots, u_{n<j}) -
\mathbb P\left (i \partial _t   + i(\alpha \cdot \nabla ) -\beta \right)(\id_4-\mathbb P)u_{j}.
\end{align}
Here, the first term on the r.h.s is already known by the inductive hypothesis and the second one is given by equation (\ref{npr}).
Hence, by induction, one can construct all higher order coefficients $u_j(t,x,\theta)$, $j\geq 1$ in this way.
\newpar
Note, that the left hand side of (\ref{prp}) is essentially a transport operator, which can be expressed as shown above.
Thus (\ref{prp}) constitutes a \emph{linear} first order system,
which determines the so-called \emph{propagating part} $\mathbb Pu_j$ from its initial data.
\newpar
\begin{remark} The above construction can be generalized to the case, where,
additionally \emph{given} external potentials $V^{ext}$, $A^{ext}$,
are included, or, equivalently, non-zero Cauchy initial data for the Maxwell equations
(\ref{dm2}), (\ref{dm3}) are assumed.
One checks that, instead of (\ref{hj}), the following Hamilton-Jacobi equation,
corresponding to $m=1$, holds:
\begin{equation}
\partial _t \phi_\pm \pm \sqrt{{|\nabla \phi_\pm - A^{ext}(t,x)|}^2+1}+ V^{ext}(t,x)=0.
\end{equation}
Since no other harmonics with $m\not=1$ exist, one again ends up
with two phases $\phi_\pm(t,x)$, corresponding to the electronic
resp. positronic degrees of freedom. In this case however, $-\phi
_+(t,x)\not = \phi_-(t,x)$, in contrast to (\ref{rel2}). Also, one
obtains an additional matrix-valued \emph{spin-transport term},
appearing on the left hand side of (\ref{trans}) and which can be
found in \cite{BK}, \cite{FK}, \cite{PST} \eg.
\end{remark}
We are now in the position to formulate our first theorem (in which we do not aim to impose the weakest possible assumptions).
In the following, $C _{(0)}^\infty$ denotes the space of smooth function, compactly supported in
$x\in \mathbb R^3$.
\begin{theorem} \label{th1}
Assume that the initial data $\psi^\varepsilon _I(x)$ admits an asymptotic expansion of the form:
\begin{equation}
\left \{
\begin{split}
& \psi_I^\varepsilon (x)=  \sqrt{\varepsilon } \, u^\varepsilon (x,\phi_I(x)/\varepsilon ),\\
& u^\varepsilon (x,\theta )\sim  \sum_{j=0}^\infty  \varepsilon^{j/2} \chi_{j}(x,\theta),
\end{split}
\right.
\end{equation}
where $\phi_I \in C^\infty(\mathbb R^3;\mathbb R)$ satisfies $\| D^2\phi_I \|\leq C$. Further, let
$\chi _j\in C _{(0)}^\infty (\mathbb R^3\times \mathbb S^1; \mathbb C^4)$ be s.t.
\begin{equation}
(\mathbb P\chi_j)(x,\theta)=\chi_j(x,\theta), \quad \forall j\in \mathbb N.
\end{equation}
Then, there exists a $0<T^*\leq T$, a corresponding domain $\Omega^*:=[0,T^*)\cap \Omega $
and a uniquely determined $u ^\varepsilon \in C^\infty _{(0)}(\Omega^*\times \mathbb S^1; \mathbb C^4)$, with
\begin{equation}
u^\varepsilon (t,x,\theta)\sim   \sqrt{\varepsilon}\, \sum_{j=0}^\infty  \varepsilon^{j/2} u_{j}(t, x,\theta),
\end{equation}
s.t. $u^\varepsilon (t,x,\phi(t,x)/\varepsilon)$ satisfies:
\begin{equation}
\left \{
\begin{split}
 i\varepsilon \partial _t u^\varepsilon - \mathcal D^\varepsilon _A (t,x,\varepsilon D) u^\varepsilon \sim & \ 0,
\quad \forall (t,x)\in \Omega^*,\\
 u^\varepsilon \big |_{t=0}= & \ \psi^\varepsilon  _I(x).
\end{split}
\right.
\end{equation}
More precisely we have:
\newpar
The principal term $u_0$ is given by (\ref{u0}),
satisfies $(\mathbb Pu_0)(t,x,\theta)=u_0(t,x,\theta)$ and
solves (\ref{teq1}) with  initial data $(\mathbb Pu_0)(0,x,\theta)= \chi_0(x,\theta)$.\\
For all $j\geq 13$, the infinite sequence of equations (\ref{npr}), (\ref{prp}), uniquely determines $u_j (t,x,\theta)$,
with initial data $(\mathbb Pu_j)(0,x,\theta)=\chi_j(x,\theta)$.
\end{theorem}
\begin{proof} The existence of a smooth phase $\phi \in C^\infty(\Omega ;\mathbb R)$, on the slab
$\Omega \subseteq \mathbb R^{1+3}$,
is already guaranteed by lemma \ref{le0}. \\
Next, consider the case $j=0$: Since $\omega_\pm(\nabla \phi )\in \mathbb R^3$, defined by (\ref{vel}), satisfies
for all multiindices $\sigma , \nu$:
\begin{equation}
\sup_{(t,x)\in\Omega}|\, \partial ^\sigma _{t}\partial _x^{\nu } \, \omega_{\pm,k}(\nabla \phi(t,x))|<\infty ,
\quad k=1,2,3,
\end{equation}
we find that the l.h.s. of (\ref{trans}) constitutes a linear symmetric hyperbolic system.
From $L^2$-conservation property (\ref{cons}) the usual commutator estimates lead to $H^s$-regularity, \ie
$u_{0} \in C^1 (\mathbb R^3; H^s)$ for all $s\geq 0$.
Now, it is a standard results for the linear wave equations in $d=3$ spatial dimensions, that
source terms in $H^s(\mathbb R^3)$ generate solutions in $H^s(\mathbb R^3)$, \cf \cite{Ho}, chapter XXIII.
This fact and Schauder's lemma imply that the maps
\begin{equation}
u_{0}(t,\cdot ) \mapsto \Gamma_\pm[u_{0}(t,\cdot )]
\end{equation}
are locally Lipschitz from $H^s( (0,t)\times \mathbb R^3 \times
\mathbb S^1)$ to itself, for all $s>2$, uniformly for $0\leq t<T$.
By a standard Picard iteration we therefore obtain a local-in-time
existence and uniqueness result in $H^s(\Omega^* \times \mathbb
S^1)$, for every $s>2$ and a Sobolev imbedding gives $u_{0}\in
C^1(\Omega^*\times \mathbb S^1; \mathbb C^4)$. The proof of the
asserted regularity for the $t$-derivatives follows by using the
differential equation to express them in terms of $x$-derivatives
and the finite speed of propagation for solution of (\ref{trans}) implies that $u_0$ is compactly supported in $\mathbb R^3_x$ since $\chi_0$ is.\\
Finally, for $j>0$, we have that the amplitudes $u_{j,\pm}$ are determined by the
linear symmetric hyperbolic system (\ref{prp}), (\ref{npr}) and the assertion is proved.
\end{proof}
Once again, we stress the fact that we analyze the MD system in a \emph{weakly coupled regime}. Indeed, the
above result implies:
\begin{corollary}
Let $u^\varepsilon(t,x,\phi(t,x)/\varepsilon)$ be as in theorem \ref{th1}, then
\begin{equation}
\left \{
\begin{split}
V^\varepsilon [u^\varepsilon](t,x) \sim  & \ \varepsilon V[u_0](t,x) + O(\varepsilon ^2),\\
A^\varepsilon [u^\varepsilon](t,x) \sim  & \ \varepsilon A_0[u_0](t,x) + O(\varepsilon ^2),
\end{split}
\right.
\end{equation}
where $V[u_0]$, $A_0[u_0]$ are nonoscillating and explicitly given by (\ref{V0}), (\ref{A0}).
\end{corollary}


\section{Stability and further results}

In theorem \ref{th1} we obtained a function $u^\varepsilon$, which solves the MD equation up to a residual $R^\varepsilon \sim 0$,
compactly supported in $[0,T^*]\times \mathbb R^3$. We want to compare $u^\varepsilon$ to a true solution $\psi^\varepsilon$ and prove
that $u^\varepsilon-\psi^\varepsilon\sim 0$ on $\Omega^*=[0,T^*]\times \mathbb R^3$.
\begin{theorem} \label{th2}
Under the assumptions of theorem \ref{th1}, there is an $\varepsilon^*\in (0,1)$, s.t. for
$\varepsilon <\varepsilon^*$, there exists a unique smooth $\psi^\varepsilon \in C_{(0)}^\infty (\Omega^*;\mathbb C^4)$,
satisfying
\begin{equation}
\left \{
\begin{split}
i\varepsilon \partial _t \psi^\varepsilon - \mathcal D^\varepsilon _A (t,x,\varepsilon D) \psi^\varepsilon = & \ 0,
\quad \forall (t,x)\in \Omega^*,\\
\psi^\varepsilon \big |_{t=0}= & \ \psi^\varepsilon  _I(x),
\end{split}
\right.
\end{equation}
which is asymptotically equivalent to $u^\varepsilon$, \ie
\begin{equation}
\psi^\varepsilon (t,x)\sim u^\varepsilon (t,x,\phi(t,x)/ \varepsilon)\quad \mbox{in $C^\infty_{(0)}(\Omega^*;\mathbb C^4)$.}
\end{equation}
\end{theorem}
\begin{proof}
Defining $v^\varepsilon:= u^\varepsilon-\psi^\varepsilon $, we obtain for the following IVP:
\begin{equation}
\left \{
\begin{split}
i(\varepsilon \partial _t + \varepsilon (\alpha \cdot \nabla)) v^\varepsilon - \beta v^\varepsilon
+ \mathcal N^\varepsilon [u^\varepsilon + v^\varepsilon ] - \mathcal N^\varepsilon [u^\varepsilon]
= & \ -R^\varepsilon ,\quad \mbox{in $\Omega^*$},\\
v^\varepsilon \big| _{t=0}= & \ 0,
\end{split}
\right.
\end{equation}
The nonlinearity can be handled analogous to the proof of lemma
6.2, in \cite{DoRa}, since for smooth sources the wave equation
has smooth solutions, which moreover travel with finite speed.
Having this in mind, the rest of the proof is a simple
modification of the one of theorem 6.1 in \cite {DoRa}.
\end{proof}
\newpar
As far as particle creation is concerned, the local-in-time solution $\psi^\varepsilon\sim O(\sqrt{\varepsilon})$
shows the following qualitative behavior:
\begin{corollary} Let $\psi^\varepsilon_I$ be as in theorem \ref{th1}. If \ $(\Pi_- \psi^\varepsilon_I)(x) = 0$,
then for $0\leq t < T^*$ it holds:
$(\Pi_- \psi^\varepsilon)(t,x) \sim O(\varepsilon ^{3/2})$, \ie no positrons are generated, up to $O(\varepsilon ^{3/2})$ and the
analogous statement for electrons is valid, too.
\end{corollary}
\begin{proof}
The assertion holds true, since a careful examination of the asymptotic expansion shows that
both, $u_0$ \emph{and} $u_1$, satisfy:
$(\mathbb P u_j)(t,x,\theta) = u_j(t,x,\theta)$, in $\Omega^*$.
\end{proof}
For completeness we shall also consider the \emph{matrix-valued Wigner transform} corresponding to $\psi^\varepsilon $, \ie
\begin{align}
w^\varepsilon [\psi^\varepsilon] (t,x,\xi):= \frac{1}{(2\pi)^3}\int_{\mathbb R^3} \psi^\varepsilon\left(t, x+\frac{\varepsilon}{2} y \right)
\otimes \overline{\psi^\varepsilon}\left(t, x-\frac{\varepsilon}{2} y\right) e^{i\xi\cdot y} dy,
\end{align}
where $\otimes$ denotes the tensor product of vectors. The $4\times4$-matrix
$w^\varepsilon [\psi^\varepsilon]$ is a \emph{phase-space description} of the quantum state $\psi^\varepsilon$.
\begin{corollary} Let $\psi^\varepsilon\sim O(\sqrt{\varepsilon})$ be the unique smooth
local-time-solution of the MD system, as guaranteed by theorem (\ref{th2}) and let
$w^\varepsilon [\psi^\varepsilon]\sim O(\varepsilon)$ be its Wigner transform.
Then, up to extraction of subsequences, we have
\begin{align}
\lim_{\varepsilon \rightarrow 0} \frac{1}{\varepsilon} w^\varepsilon [ \psi^\varepsilon] = \mu
\quad\mbox{in $\mathcal S'([0,T^*)\times \mathbb R_x^3\times \mathbb R_{\xi}^3)$ weak-$\star$,}
\end{align}
where the matrix-valued Wigner measure $\mu$ is given by: $\mu =\mu_+ + \mu_-$, with
\begin{align}
\mu_\pm(t,x,\xi)= u_{0,\pm}(t,x)\otimes \overline {u}_{0,\pm}(t,x) \, \delta (\xi \mp \nabla \phi(t,x)).
\end{align}
\end{corollary}
\begin{proof}
Since $\phi$ has no stationary points within $\Omega^*$, a nonstationary phase argument
implies that all Wigner matrix elements, which mix the electronic and positronic components are
of order $O(\varepsilon ^{\infty})$. The assertion then follows from the
well known results on Wigner measures, \cf \cite{GMMP}.
\end{proof}
We finally remark on the case of the Dirac-Maxwell system where the Dirac particles have vanishing mass.
Instead of (\ref{hj}) we obtain
\begin{equation}
\partial_t \phi= \pm |\nabla \phi| ,
\end{equation}
which is equivalent to the eikonal equation of the wave equation.
It follows that in this case lemma \ref{le1} can not hold, since
the phases $\pm \phi$ are characteristic for the wave equation.
More precisely, they are indeed \emph{everywhere} characteristic,
\ie in all of $\Omega$, which again allows for an asymptotic
description of the $A^\varepsilon$, similar to (\ref{aasy}),
(\ref{A1}), \cf \cite{La} or \cite{Ra}, chapter 5. In this case,
the $\varepsilon$-oscillations are also given by $\exp(\pm 2i\phi
/ \varepsilon)$, but the corresponding amplitudes $A_l^\pm$ are of
course different. The main difference, however, is the fact that
in this case the summation index runs from $l=0$ to infinity, \ie
$\varepsilon$-oscillations are present already in the lowest order
term. This leads to a more complicated structure of the transport
equations for the amplitudes $u_j$, but apart from that all
results remain valid.
\vskip 1 cm
\textbf{Acknowledgement:}
\newpar
This work originated from a discussion with P. G\'erard (Univ. d\'e
Paris Sud). The authors thank J. Rauch and H. Spohn for helpful
discussions and the \emph{TICAM} (Univ. Texas) for its support. This work
has also been supported by the Austrian Science Foundation FWF
through grant no. W8, project no. P14876-No4 and the Wittgenstein
Award 2000 of P. M. Additional financial sponsorship has been
given by the European Union research network \emph{HYKE}.



\end{document}